# Molecular Dynamics Simulations Indicate that Deoxyhemoglobin, Oxyhemoglobin, Carboxyhemoglobin, and Glycated Hemoglobin under Compression and Shear Exhibit an Anisotropic Mechanical Behavior

Sumith Yesudasan, Xianqiao Wang, and Rodney D. Averett*

*School of Chemical, Materials, and Biomedical Engineering, College of Engineering, University of Georgia, 597 D.W. Brooks Drive, Athens, GA 30602*

*Corresponding author email: raverett@uga.edu
Telephone: 706-542-0863

## Abstract

We developed a new mechanical model for determining the compression and shear mechanical behavior of four different hemoglobin structures. Previous studies on hemoglobin structures have focused primarily on overall mechanical behavior; however, this study investigates the mechanical behavior of hemoglobin, a major constituent of red blood cells (RBCs), using steered molecular dynamics (SMD) simulations to obtain anisotropic mechanical behavior under compression and shear loading conditions. Four different configurations of hemoglobin molecules were considered: deoxyhemoglobin (deoxyHb), oxyhemoglobin (HbO$_2$), carboxyhemoglobin (HbCO), and glycated hemoglobin (HbA$_{1C}$). The SMD simulations were performed on the hemoglobin variants to estimate their unidirectional stiffness and shear stiffness. Although hemoglobin is structurally denoted as a globular protein due to its spherical shape and secondary structure, our simulation results show a significant variation in the mechanical strength in different directions (anisotropy) and also a strength variation among the four different hemoglobin configurations studied. The glycated hemoglobin molecule possesses an overall higher compressive mechanical stiffness and shear stiffness when compared to deoxyhemoglobin, oxyhemoglobin, and carboxyhemoglobin molecules. Further results from the models indicate that the hemoglobin structures studied possess a soft outer shell and a stiff core based on stiffness.

## Keywords

## I. Introduction

Understanding the molecular mechanical properties of thrombi (Weisel, 2004) and their constituents (Aleman, Walton, Byrnes, & Wolberg, 2014; Lai, Zou, Yang, Yu, & Kizhakkedathu, 2014; Loiacono et al., 1992; Pretorius & Lipinski, 2013; van der Spuy & Pretorius, 2013; van Gelder, Nair, & Dhall, 1996; Wang et al., 2016; Adam R. Wufsus et al., 2015) is necessary for understanding the bulk mechanical and physiological function of the thrombus. Fibrin clots consist mainly of fibrin (precursor is the molecule fibrinogen), platelets, and erythrocytes. The mechanical strength of fibrinogen has been studied in the past years using experiments (Brown, Litvinov, Discher, & Weisel, 2007; Carlisle et al., 2009; Gottumukkala, Sharma, & Philip, 1999; McManus et al., 2006; Weisel, 2004), simulations (Gubskaya, Kholodovych, Knight, Kohn, & Welsh, 2007; Isralewitz, Gao, & Schulten, 2001; Lim, Lee, Sotomayor, & Schulten, 2008), and multiscale approaches (Govindarajan, Rakesh, Reifman, & Mitrophanov, 2016; Perdikaris, Grinberg, & Karniadakis, 2016; Piebalgs & Xu, 2015). Recent years have witnessed various experimental studies on the estimation of the mechanical properties of fibrin clots (Foley, Butenas, Mann, & Brummel-Ziedins, 2012; Riha, Wang, Liao, & Stoltz, 1999; Tocantins, 1936; Weisel, 2004). For example, the estimation of the mechanical properties of bulk thrombi (Krasokha et al., 2010) and cross-linked fibrin networks (Weisel, 2004) has been reported. The mechanical behavior of a single fibrin fiber was also studied (Liu et al., 2006) by stretching the fibrin fibers using atomic force microscopy (AFM) tip and fluorescent microscopy, which reports that the extensibility of fibrin films is in the range of 100-200% and for single fibrin fibers it is estimated at 330%. In addition, the physiological path and functional steps in fibrin clot formation has been discussed in numerous research studies (Cito, Mazzeo, & Badimon, 2013; Undas & Ariëns, 2011; A. Wufsus, Macera, & Neeves, 2013).

Despite the vast literature in the physiological understanding and mechanical modeling domain of fibrin clots, the role of the mechanical strength of constituents such as RBCs are seldom discussed or poorly understood from a molecular basis. Thus, understanding the mechanical properties of the allosteric protein hemoglobin (a major constituent of RBCs) is important for the development of advanced mechanical models of thrombi (clots) with inclusions (Kamada, Imai, Nakamura, Ishikawa, & Yamaguchi, 2012; Loiacono et al., 1992; Mori et al., 2008; Wagner, Steffen, & Svetina, 2013).

It is well established that RBCs (inclusions of thrombi) must withstand the increasing pressure of blood flow and other forces (Svetina, Kuzman, Waugh, Ziherl, & Žekš, 2004; Teng et al., 2012; Uyuklu, Meiselman, & Baskurt, 2009; Wu, Guo, Ma, & Feng, 2015; Yoon & You, 2016), as this may lead to plastic deformation of the thrombus and eventually rupture (Weisel, 2004). The focus of these studies, however, has been primarily at the continuum level without much focus on molecular mechanical behavior. Some researchers have performed investigations on the mechanical and dynamic behavior of hemoglobin from a molecular standpoint (Arroyo-Mañez et al., 2011; Kakar, Hoffman, Storz, Fabian, & Hargrove, 2010; Koshiyama & Wada, 2011; Xu, Tobi, & Bahar, 2003), and some studies that have been focused on the molecular mechanical behavior of hemoglobin variants such as sickled hemoglobin (HbS) (Li, Ha, & Lykotrafitis, 2012; Li & Lykotrafitis, 2011) and glycated hemoglobin (De Rosa et al., 1998). There have also been studies that have addressed the anisotropic behavior of hemoglobin structures using fluorescence based methods (Bucci & Steiner, 1988; Chaudhuri, Chakraborty, & Sengupta, 2011; Hegde, Sandhya, & Seetharamappa, 2013; Kantar, Giorgi, Curatola, & Fiorini, 1992; Plášek, Čermáková, & Jarolím, 1988) but did not address the molecular mechanical behavior. In sum, these previous investigations were limited since they did not explore the molecular mechanical behavior of hemoglobin (or its variants) under mechanical compression or shear loading conditions. Because biological cells (in particular RBCs) experience compressive and shear forces physiologically and must exhibit appropriate transport behavior, an investigation that explores the molecular anisotropic mechanical behavior of the comprising proteins under these loading conditions would be a significant enhancement to the scientific literature.

In this work, we investigated the mechanical strength of various forms of hemoglobin, a major component of RBCs, from an atomistic viewpoint, utilizing steered molecular dynamics (SMD) simulations. Four different types of hemoglobin molecules (deoxyhemoglobin, oxyhemoglobin, carboxyhemoglobin, and glycated hemoglobin) were considered to estimate their unidirectional stiffness and shear stiffness at the molecular level. The results of this work are important for the development of advanced cellular mechanical models in biophysics and bioengineering.

## II. Methods
### A. Molecular Dynamics Model

Hemoglobin (Hb) is a molecule that is considered a globular protein consisting of four subunits (Fig. 1a). Two of these subunits represent the $\alpha$ chains and the other two represent the $\beta$ chains. Each of these subunits consists of a *heme* group with an iron atom in the center (shown as bonded molecular representation in Fig. 1a). Here we consider four different types of Hb molecules, namely deoxyhemoglobin (RCSB 2DN2), oxyhemoglobin (RCSB 2DN1), carboxyhemoglobin (RCSB 2DN3), and glycated hemoglobin (RCSB 3B75). The deoxyhemoglobin molecule (deoxyHb) is the form of Hb without any additional molecules, the oxyhemoglobin molecule (HbO$_2$) represents the oxygen carrying state of Hb, and the carboxyhemoglobin molecule (HbCO) represents the carbon monoxide structure attached to the Hb molecule. The glycated Hb structure represents the Hb molecule with bonded glucose and fructose molecules. These Hb *.pdb models were solvated in a spherical water volume (Fig. 1b) and used for molecular compressive and shear stiffness calculations. The molecular models were prepared with Visual Molecular Dynamics (VMD) (Humphrey, Dalke, & Schulten, 1996) and the molecular dynamics (MD) simulations are performed with NAMD (Kalé et al., 1999). The force field employed for the MD simulations was CHARMM27 (MacKerell, Banavali, & Foloppe, 2000) and the water molecular model used for solvation was TIP3P (Jorgensen, Chandrasekhar, Madura, Impey, & Klein, 1983). For equilibration, an energy minimization was performed for 1,000 steps and then equilibrated using a Langevin thermostat at 300 K for 10,000 steps followed by an NVE relaxation for 40,000 steps. This equilibrated model was used for production runs (compression studies) with controlled temperature using a Langevin thermostat (Allen & Tildesley, 1989).

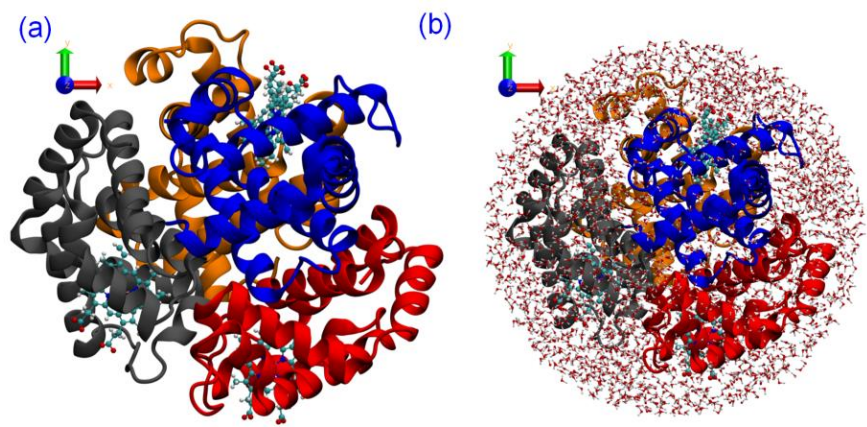

**Figure 1.** Computational models of human hemoglobin. (a) Deoxyhemoglobin (deoxygenated) in the absence of solvent. (b) Hb solvated in a sphere.

## B. Alignment of models

The preliminary molecular models (PDB files) were arbitrarily oriented in the Cartesian coordinate system. In order to make reliable comparisons of the unidirectional stiffness and other mechanical properties among these four configurations, a consistent and geometrically similar arrangement of these four molecular models was necessary. A geometric basis for all four Hb configurations was developed using the iron atoms of the *heme* group, and they were consistently aligned using a two-step rotation transformation process (Fig. 2). These aligned Hb molecule configurations were used for all SMD analysis studies performed in this work.

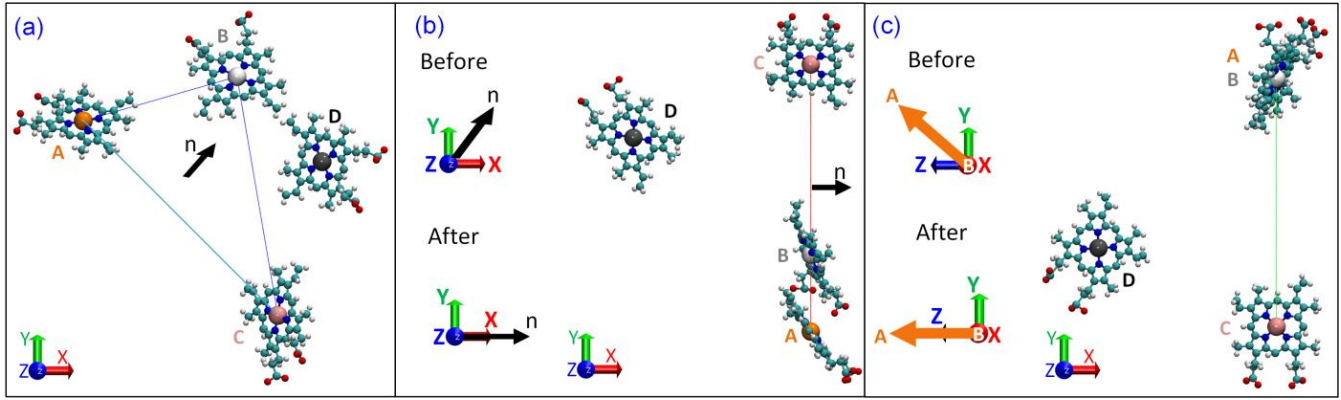

**Figure 2:** Alignment of the Hb molecules. (a) Sample initial orientation of the Hb molecule, with heme group only. A, B, C and D represent the different subunits and *n* is the normal vector of triangle ABC. (b) The Hb structure was rotated by aligning the normal of ABC with the x-axis. (c) The edge AB of the triangle ABC was then aligned with the z-axis. All views of the molecules shown here are from the xy-plane.

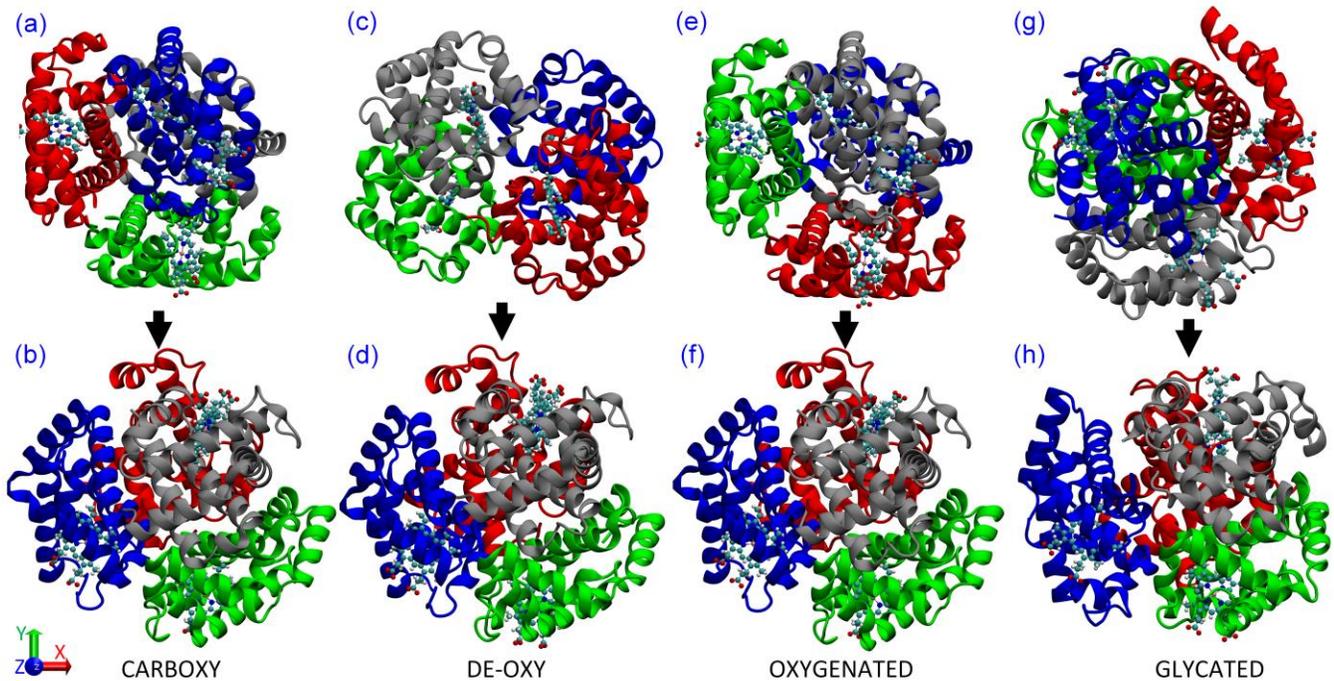

**Figure 3:** Aligned Hb molecules. The initial and rotated molecular models of hemoglobin are shown for (a-b) carboxyhemoglobin, (c-d) deoxyhemoglobin, (e-f) oxyhemoglobin and (g-h) glycated hemoglobin, respectively.

The iron atoms of heme of different subunits were constructed using VMD (A, B, C and D, Fig. 2a). Consider the triangle formed by A, B and C. Let *n* be the normal of this triangle ABC. For consistent orientation arrangement, as a first step we shifted the Hb molecule by corner A to the origin and performed a rotation to align normal *n* with the x-axis (Fig. 2b). Next, edge AB was aligned with the z-axis by a single rotation (Fig. 2c). This translation and rotation procedure was used for all four Hb molecule variants, namely carboxy, deoxy, -oxy and glycated hemoglobin (Fig. 3a-b, Fig. 3c-d, Fig. 3e-f, and Fig. 3g-h, respectively). Geometric similarity was observed for the aligned molecules, with the exception of glycated Hb. The glycated Hb molecule possesses a different internal structural due to the presence of embedded glucose and fructose molecules.

## C. Stiffness Estimation studies

In this study, stiffness of the molecule is defined as the force required for unit deflection. The unidirectional stiffness was estimated by compressing the system from either side along the x, y and z directions. The shear stiffness was estimated by applying opposing directional tangential forces on either side of the Hb molecule.

## 1. Unidirectional stiffness

Forces were applied on the peripheral atoms of the spherically solvated Hb molecule, which mimic compression using two imaginary rigid plates. A graphical representation of the force application strategy is shown in Fig. 4a. Two imaginary rigid plates move towards the center of the Hb molecule from both sides. These plates were initially kept at a distance $d_0$ from the center of the Hb molecule, which was also the origin of the coordinate system. During the SMD simulation, the imaginary plates were moved towards the center at a constant velocity $v$. At any instantaneous time ($t$) from the beginning of the simulation, the location of the plates was given as: $d_0 - vt$. From this location, a region of influence was defined at distance $r_c$. A force ($f_i$) was applied to the $i^{th}$ atom which occupies the region: $|\vec{r}_i.\vec{n}| > d_0 - vt - r_c$. The applied forces on atoms are distance dependent and quadratic in nature, to ensure the Hb molecules were sufficiently compressed, and to mimic a rigid wall. The force assumes the form of a quadratic curve initiating from zero at the point ($d_0 - vt - r_c$) of influence (Fig. 4b), gradually increasing towards the imaginary plate. The gradient on the plate (Fig. 4a) depicts the magnitude of the force applied to the molecule, small near $d_0 - vt - r_c$ and significantly large near $d_0 - vt$. The force was applied only to the qualifying atoms of Hb at every 1 ps. This means, at every 10$^{th}$ step of time integration, the external force application is turn on, which simulates a gradual compression instead of shock force application. A TCL script in conjunction with NAMD was used to impose these force criteria onto the system.

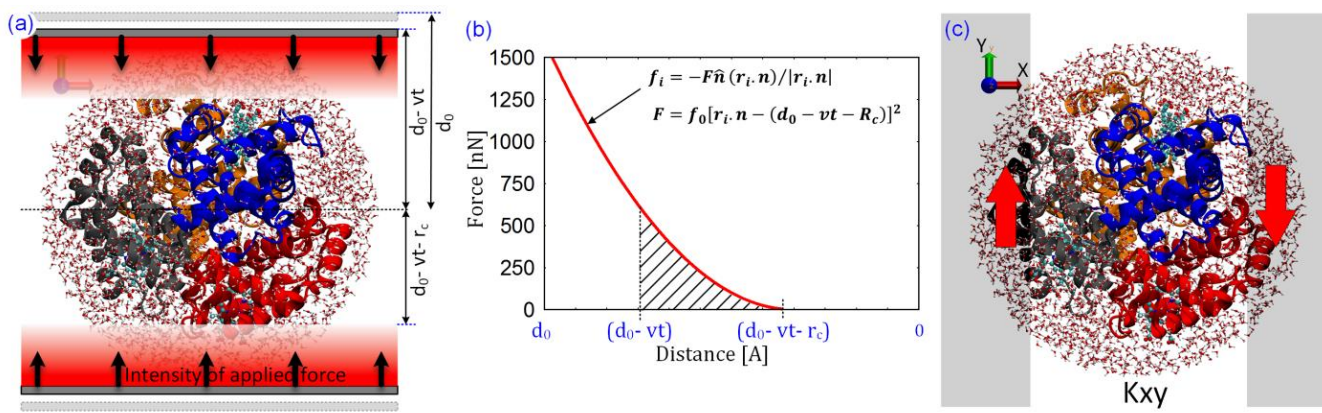

**Figure 4:** Compression strategy for Hb structures. (a) Two imaginary rigid plates travel at a constant velocity towards the center of the molecule and exerts a variable force on the atoms. The magnitude of the force applied is depicted as the gradient intensity (red). (b) Applied force vs. distance, where the horizontal axis represents the distance from the origin. The force of influence in atoms at a particular time *t* is shown as a shaded diagonal pattern. (c) The atoms occupying the shaded region were selected and an applied force along the directions was used to simulate shear mechanical behavior.

Utilizing Equations (1) – (2), atoms occupying the qualifying region ($|\vec{r}_i.\vec{n}| > d_0 - vt - r_c$), were used to compute and apply the requisite mechanical forces.

$$\vec{f}_i = -F\vec{n}\frac{\vec{r}_i.\vec{n}}{|\vec{r}_i.\vec{n}|} \quad (1)$$

$$F = f_0\left[\vec{r}_i.\vec{n} - (d_0 - vt - r_c)\right]^2 \quad (2)$$

Here, $n$ is the normal vector to the imaginary plates, $r_i$ is the position vector of $i^{th}$ atom, $f_0 = 100$ pN, $v = 1$ A/ps, and $r_c = 0.8$ nm. This study was difficult to accomplish in a periodic boundary due to the pressure fluctuations arising from the localized density variations and probability of cavitation.

**2. Shear stiffness**

Shear stiffness is defined as the shear force required for unit deflection along the shear stress direction. In the case of shear, estimation of six shear components corresponding to *xy, xz, yx, yz, zx,* and *zy*, tensor components was necessary. For example, shear stiffness in the *xy* direction is defined as $k_{xy} = F_{xy}/d_{xy}$, where $k_{xy}$ is the shear stiffness, $F_{xy}$ is the force acting on the shear plane *x*, along the *y*-direction, and $d_{xy}$ is the deflection along the *y*-direction. In general, for any shear plane with unit normal $n_1$, the shear stiffness along unit normal direction $n_2$ was defined as $k n_1 n_2 = F n_1 n_2 / d n_1 n_2$. The shear force was applied to the atoms of the Hb molecule using two criteria: 1) an atom selection was performed based on Equation (3) and 2) the force on the $i_{th}$ atom was defined by Equation (4). A graphical explanation of shear force application and selection of the atoms is described for the case of shear stiffness calculation in the *xy*-plane (Fig. 4c).

$$|(\vec{r}_i - \vec{r}_{com}).\vec{n}_1| > d_0 - r_c \quad (3)$$

$$\vec{f}_i = -f_0\vec{n}_2\frac{\vec{r}_i.\vec{n}_1}{|\vec{r}_i.\vec{n}_1|} \quad (4)$$

Equation (3) is a selection criterion, from which the qualifying atoms of the Hb molecule within the influence region of the imaginary rigid surface were selected and individually applied with a force $f_i$. For this study analysis: $d_0 = 3$ nm, $r_c = 1.5$ nm, $f_0 = 100$ pN, and $r_{com}$ is the position vector of the center of mass of the Hb molecule, which is (0,0,0) in our case.

## III. Results and Discussion
### A. Unidirectional Stiffness

The rotated and aligned models of hemoglobin were used to create two molecular models: 1) spherically solvated model in water and 2) a plain model in the absence of water (non-solvated). These two computational models were then used to compress Hb along the three directions *x*, *y* and *z* individually using the force application method explained in the aforementioned. The SMD simulation was performed for 40 ps, with a time step of 1 fs, and the compression algorithm was applied using a TCL script at every 0.01 ps. Evolutionary mechanical behavior of deoxyhemoglobin at 10 ps intervals during x-axis compression (Fig. 5) indicates that 1) the Hb molecule is compressed tightly into a discoid structure (front view) 2) experiences a gradual separation of the subunits (side view), and 3) experiences separation of the various amino acid residues (coiled-coil regions).

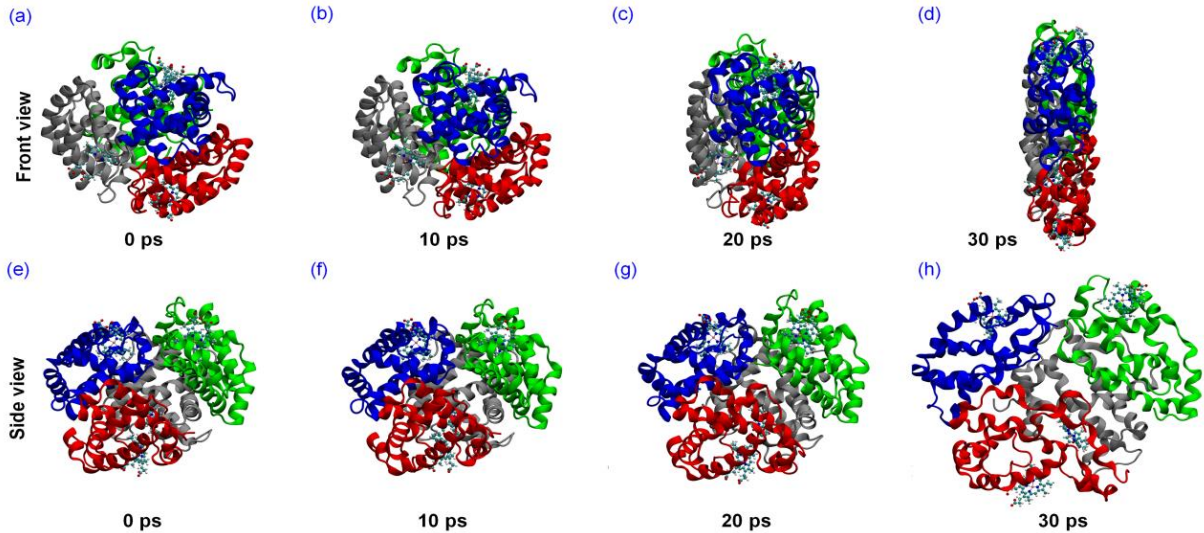

**Figure 5:** Uniaxial compression of deoxyhemoglobin molecules. Front view of deoxyHb during compression at (a) initially, (b) 10 ps, (c) 20 ps and (d) 30 ps. Side view of system under compression at (e) 0 ps, (f) 10 ps, (g) 20 ps, and (h) 30 ps.

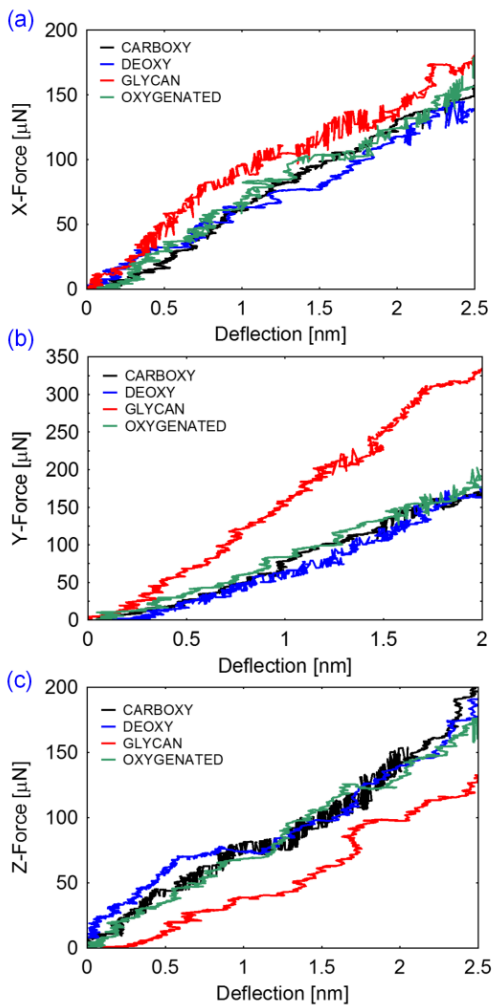

**Figure 6:** Force versus deflection comparison for various hemoglobin structures: deoxyhemoglobin (deoxyHb), oxyhemoglobin (HbO$_2$), carboxyhemoglobin (HbCO), and glycated hemoglobin (HbA$_{1C}$). The directional forces and corresponding deflections are shown for (a) x-axis, (b) y-axis, and (c) z-axis respectively.

The total force applied to the system was recorded every 1 ps and the deflection was estimated from the trajectories of the atoms in the system. The force versus deflection behavior along all three directions for carboxy, deoxy, -oxy and glycated hemoglobin molecules is shown in Figs. 6a, 6b, and 6c respectively. The slope estimated from the curves in Fig. 6 provides the unidirectional stiffness of Hb molecules.

The slopes (unidirectional stiffness) are shown in the Fig. 7a for both solvated and non-solvated cases. The overall trend shows a more rigid glycated Hb structure compared with the other Hb variants. In addition, the stiffness along the y-axis shows almost twice the magnitude in the other two directions *x* and *z*, which reveals an anisotropic material property of hemoglobin. This anisotropic mechanical behavior was observed in all Hb variants.

## B. Shear Stiffness

The shear force on the Hb molecules was applied as per the strategy in the aforementioned, for all the 6 possible directions, namely *xy*, *xz*, *yx*, *yz*, *zx*, and *zy*. MD simulations were performed for every case with a 10 ps duration and a time integration step of 1 fs. The shear force algorithm was applied using a TCL script coupled with NAMD, at every 10 fs. The width of the influence plane used to select atoms (Fig. 4c) is considered as 1.5 nm on either side.

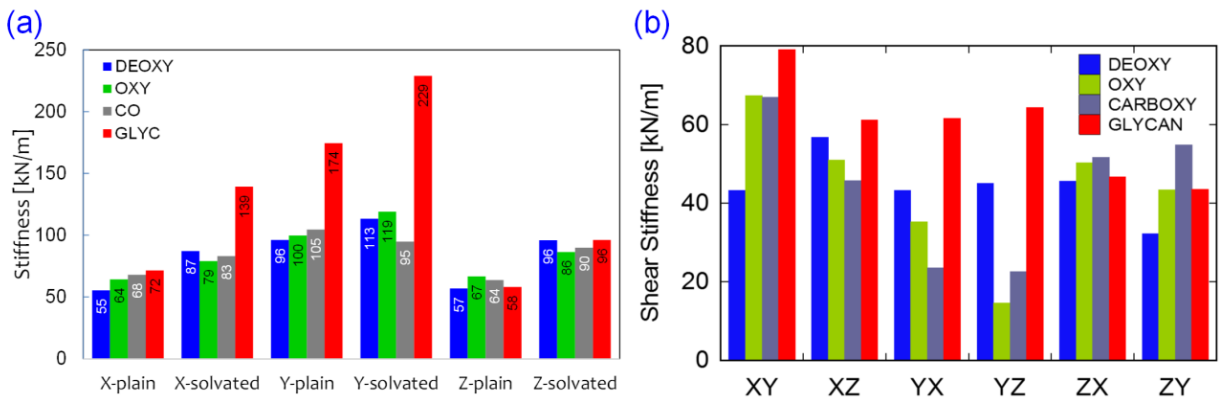

**Figure 7:** Stiffness comparison for various hemoglobin variants. (a) Unidirectional stiffness values of deoxy, oxy, carboxy, and glycated hemoglobin structures are shown for solvated and non-solvated cases. (b) Shear stiffness along XY, XZ, YX, YZ, ZX and ZY planes for the four hemoglobin variants.

The shear stiffness was estimated for non-solvated (plain) models (Fig. 7b). In most of the configurations, the glycated Hb molecule exhibits a higher shear stiffness. The shear stiffness of the glycated Hb is much higher than the other Hb molecules for the cases *yx* and *yz*. The shear stiffness of glycated Hb in the z-plane along y and x direction shows a lower or equal stiffness as compared with the other Hb variants, which is consistent with the unidirectional results.

To verify the sensitivity of the applied force magnitudes ($f_0$) and compression rate ($v$), we performed a set of separate simulations and found that there is no significant effect on the stiffness values of Hb. For all three directions (*x*, *y*, and *z*), the solvated models show higher strength than the non-solvated models. The results also show an augmented stiffness of the Hb molecules in the presence of water as a solvent. The glycated hemoglobin possesses almost twice the stiffness of other configurations.

## C. Energy Change and RMSD

To understand the molecular mechanism and origin of the stiffness, we have investigated the various energy contributions and the root mean square deviation (RMSD) response of the four Hb molecules during compression. The compressive strength of the molecule is defined as the ability to resist deformation, which becomes the ability to resist the change in potential energy at an atomistic level. Therefore, the estimation of change in potential energy during compression may shed light into the origin of the mechanical strength of the hemoglobin structures. The energy change of the hemoglobin systems during compression along the *x*, *y*, and *z* directions was computed (Fig. 8a, 8b, and 8c respectively). The estimated energy was averaged with the total number of atoms in the system (kcal/mol). The energies arising from the potentials including bond (BOND), angle (ANG), dihedral (DIHED), improper (IMPR), coulomb (ELEC), and van der Waals (VDW) were calculated (Fig. 8). The kinetic energy (KE), total potential energy (PE) and total energy (TOTAL) was also computed (Fig. 8).

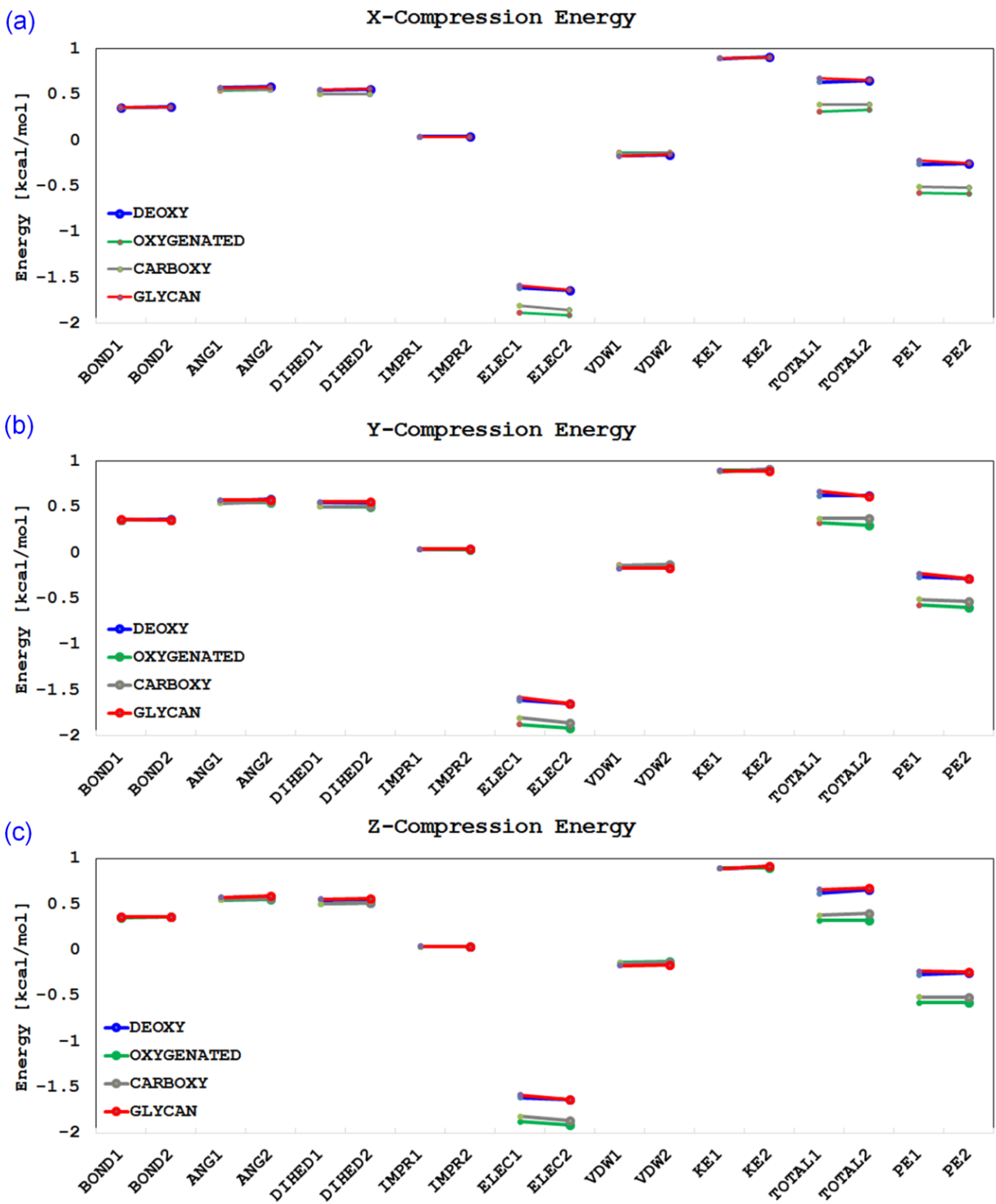

**Figure 8:** Energy of the hemoglobin system before and after compression. The energy corresponding to (a) x-axis compression, (b) y-axis compression, and (c) z-axis compression (1 and 2 correspond to before and after compression, respectively). The energy is plotted for the four hemoglobin variants.

As displayed in Fig. 8, the results do not show much variation in the energy before (suffix 1) and after (suffix 2) compression. The major contribution to the potential energy was Coulombic energy and the least contribution was from improper potential energy. A more in-depth analysis of the energy change during compression in terms of percentage change was calculated using: $\Delta E = 100(E_1 - E_2)/E_1$ (Figure 9). From this percentage change, the glycated Hb molecule shows relatively large changes in potential energy during compression along the x-axis (10%) and y-axis (24%). When compressed along the z-axis, there are not considerable variations in energy across various Hb configurations. These potential energy changes follow the same trend as the unidirectional stiffness of the Hb molecules and hence they are directly related.

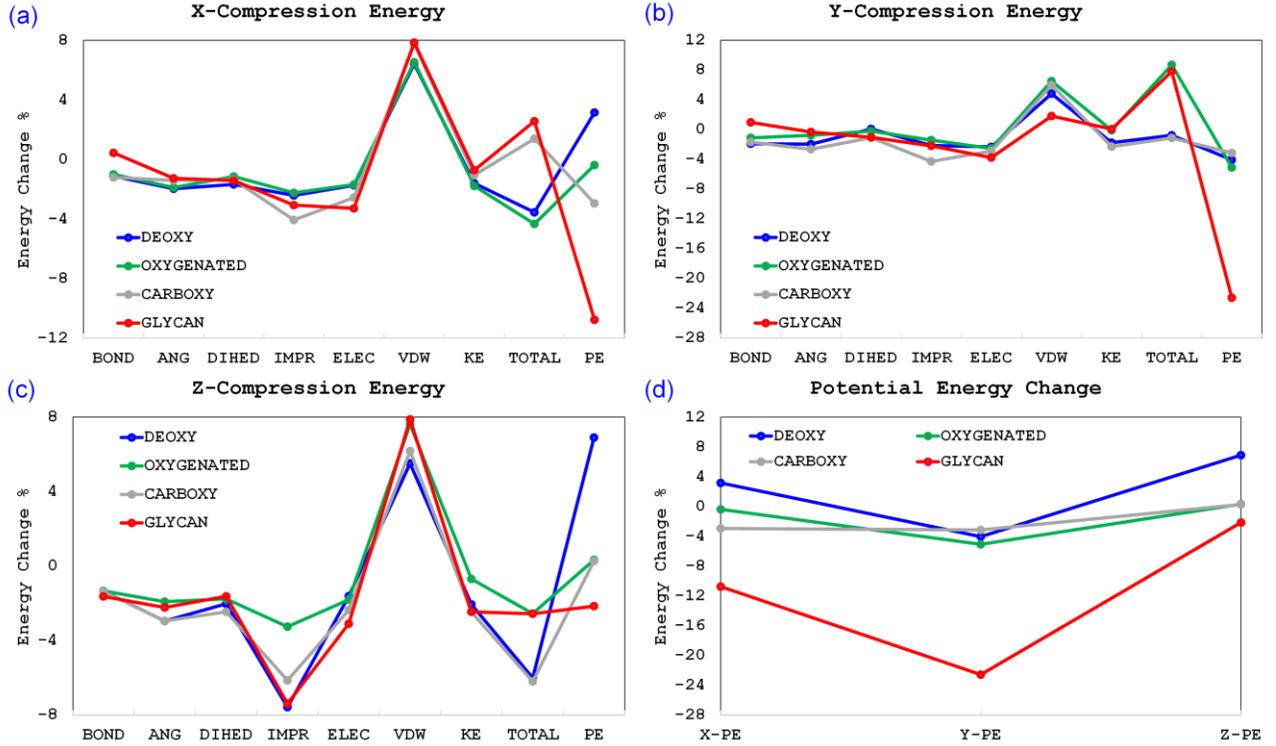

**Figure 9:** Percentage energy change for the four hemoglobin variants in (a) x-direction, (b) y-direction and (c) z-direction. (d) Percentage change in potential energy for all hemoglobin structures in x, y, and z directions. The variations closely follow the trend in stiffness property.

Root mean square deviation (RMSD) of the 4 iron atoms present in the heme residues of the α chains and the β chains of the Hb molecules during compression were computed. The RMSD of these iron atoms while compressing the Hb molecules along x, y and z directions respectively is displayed (Fig. 10a, 10b, and 10c). The RMSD was estimated based on Equation 5. Equation 5 estimates the RMSD between present and previous states of the ensemble of atoms, where $x$, $y$ and $z$ are the positions of the $N$ atoms, and subscripts $i$ and $t$ represents $i$-th atom and time $t$, and $\Delta t$ represents the time step.

$$RMSD = \sqrt{\sum_{i=1}^{N} \begin{pmatrix} (x_{it} - x_{i(t-\Delta t)})^2 + \\ (y_{it} - y_{i(t-\Delta t)})^2 + \\ (z_{it} - z_{i(t-\Delta t)})^2 \end{pmatrix} / N} \quad (5)$$

The maximum RMSD of the iron atoms for all the cases was also calculated (Fig. 10d).

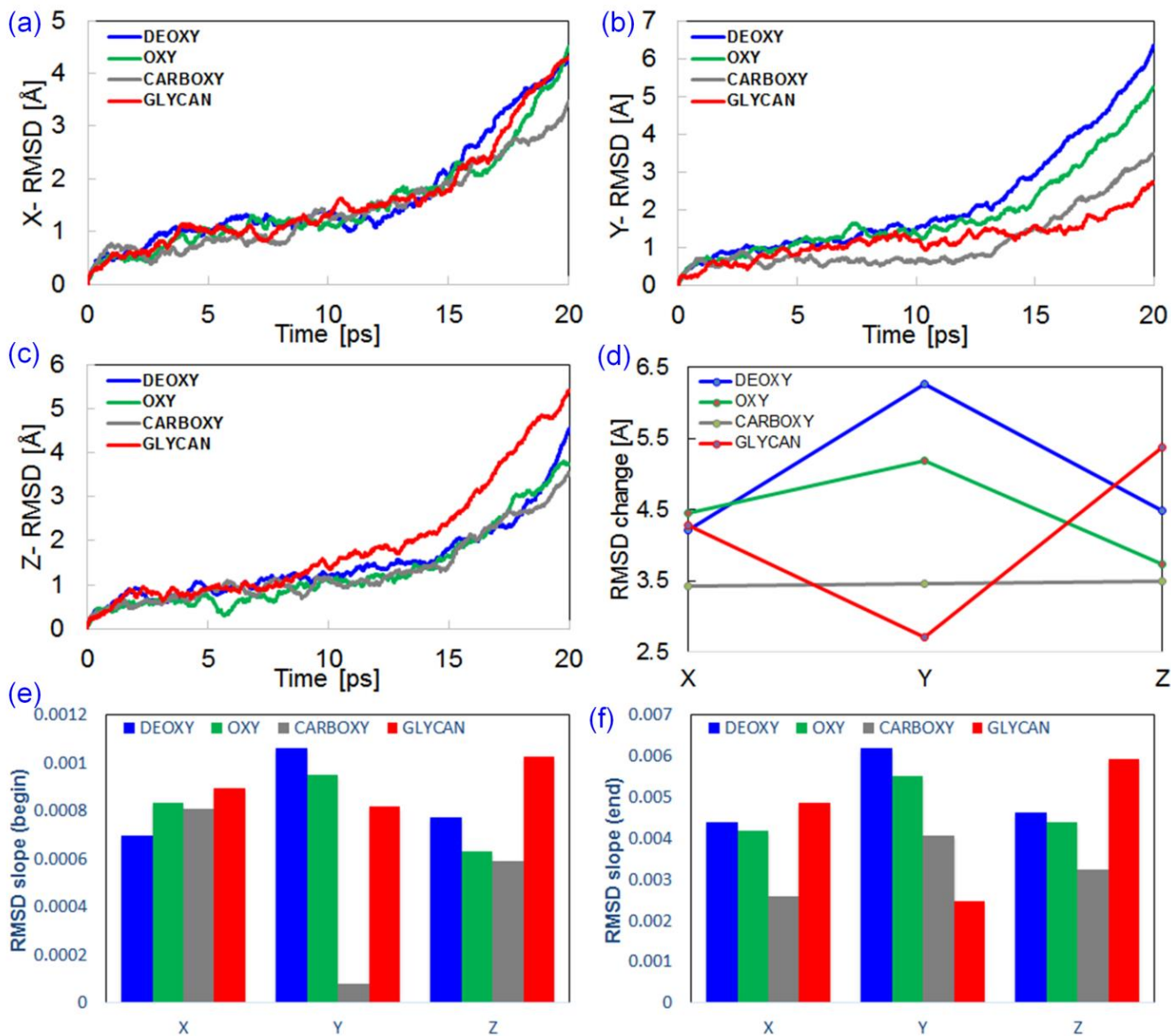

**Figure 10:** RMSD evolution of Fe atoms of heme residue for first 20 ps during (a) x-direction compression, (b) y-direction compression, and (c) z-direction compression. The slope of this RMSD curve was estimated for (e) initial 3 ps and (f) final 3 ps and plotted. This shows the rate of compression. (d) The final RMSD of the hemoglobins in all directions are shown.

The trend of the RMSD shows a linear slope in the beginning of the compression (0 ps to 5 ps) and a steep nonlinear mechanical behavior towards the end of compression (15 ps to 20 ps) (Fig. 10a-c). These slopes were evaluated separately (Fig. 10e and 10f). The RMSD slope mathematically correlates to the expansion rate of the Fe atoms, and increases by a factor of 6 towards the end of compression. Since the RMSD is dependent on the atomic coordinates on all three directions $x$, $y$ and $z$, the results in Fig. 10 show that the Fe atoms approach one another during compression along the loading axis at the initiation. When the Hb molecules compress and flatten like an oval disc (Figs. 5d and 5h), these Fe atoms separate perpendicular to the compression axis. This also shows that the core molecules in the heme residue actively participate in the structural activity after the 'shell' or peripheral molecules become compressed.

## D. Solvation Volume Sensitivity and Hydrogen Bonds

A cutoff potential of 1 nm was used for all MD simulations in this work with a water solvation volume of 1.5 nm thickness surrounding the Hb molecules. The effects of larger solvation spheres on stiffness calculations are seldom due to the non-periodic boundary condition or the short-range potential cutoff of 1 nm.

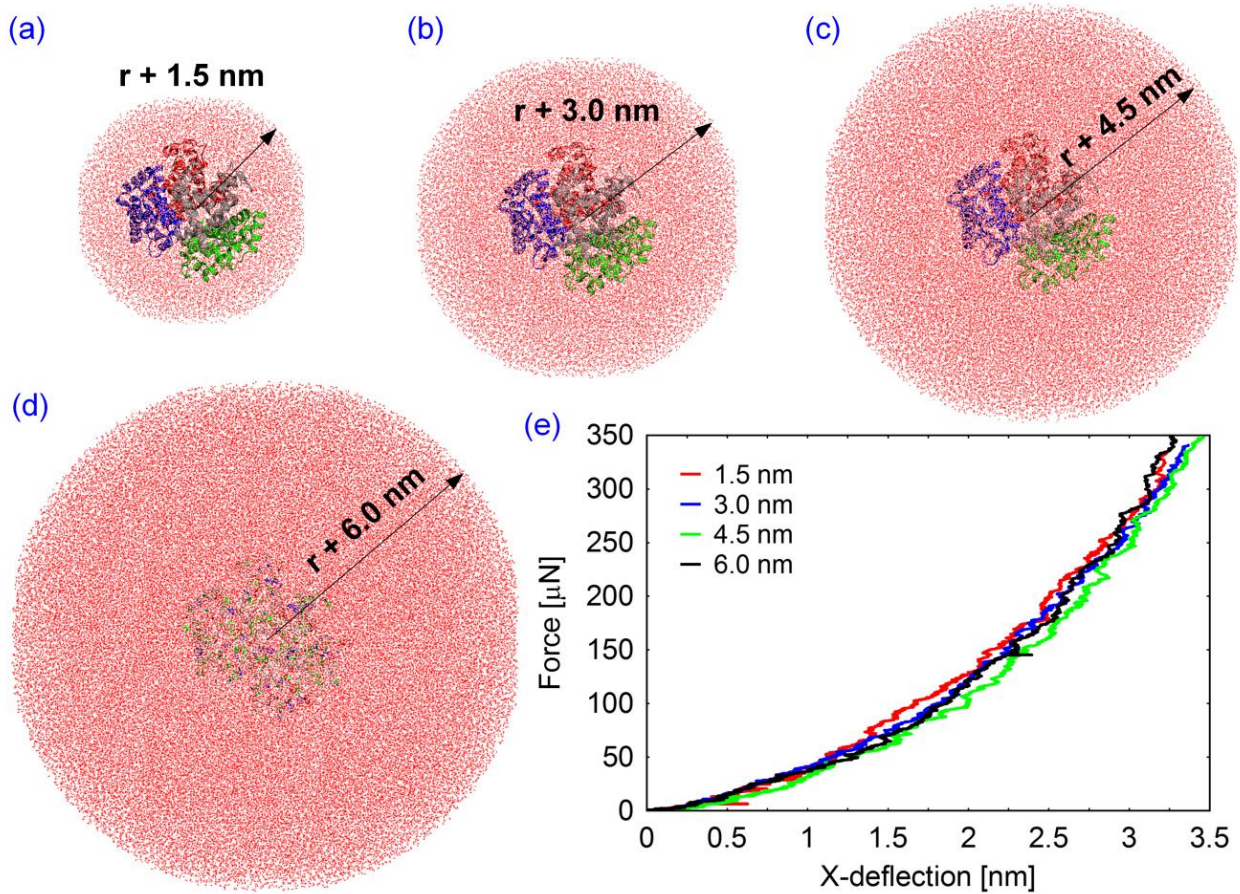

**Figure 11.** Sensitivity study of solvation volume influence on stiffness of the Hb molecule. Solvated models of hemoglobin with water envelope of (a) 1.5 nm, (b) 3 nm, (c) 4.5 nm and (d) 6 nm. (e) Force vs. deflection graph for x-direction compression for oxyhemoglobin with different solvation volumes, where $r$ is the radius of the Hb molecule.

To corroborate this non-sensitivity, we solvated the Hb molecules in various spheres ranging from 1.5 nm to 6 nm envelopes (Fig. 11a-d). The deflection of the Hb molecules was then computed by applying compressive forces along the x-direction. The compression results (Fig. 11e) do not display any influence of bigger solvation spheres.

To understand the possible role of hydrogen bonds (H-bonds) on the stiffness of the Hb molecules, the number of bonds formed was calculated during the compression process using VMD (Humphrey et al., 1996). Figure 12 shows the number of hydrogen bonds formed between Hb molecules and water during the compression process for the various Hb molecules along different directions. For all four Hb variants, the calculations show no change in the number of hydrogen bonds formed. Subsequent to 15 ps, there is a steady decline in the number of hydrogen bonds and this is correlated with the RMSD results (Fig. 10a-c), where a steady compression is observed after 15 ps. These results indicate that hydrogen bonds contribute significantly to Hb compressive stiffness, and in particular explain the augmented stiffness in the solvated case.

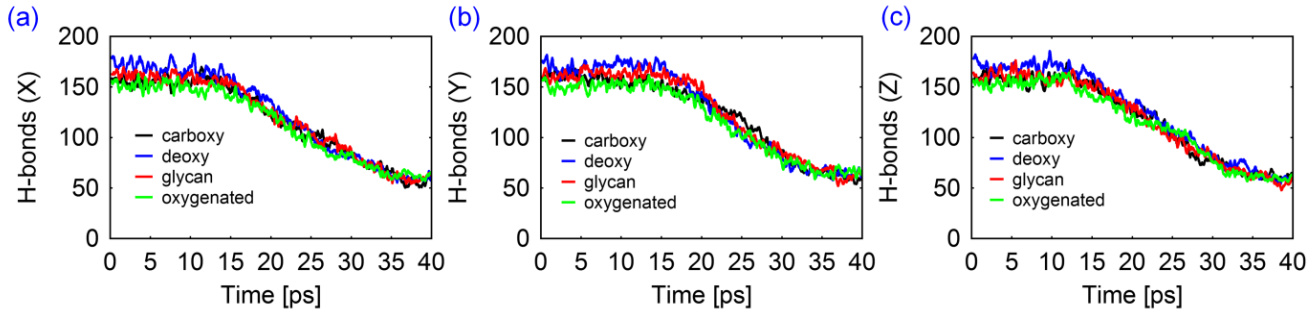

**Figure 12:** Evolution of the number of hydrogen bonds formed during compression along (a) x-axis compression, (b) y-axis compression and (c) z-axis compression.

## IV. Conclusions

In this paper, we have investigated the mechanical properties of various forms of hemoglobin in different physiological states. Unidirectional stiffness and shear stiffness were calculated for deoxyhemoglobin (deoxyHb), oxyhemoglobin (HbO$_2$), carboxyhemoglobin (HbCO), and glycated hemoglobin (HbA$_{1C}$). The unidirectional stiffness varied significantly among these four configurations, where the glycated Hb displayed the highest stiffness compared to the other three hemoglobin variants. With respect to shear stiffness, the results show a similar trend to that of unidirectional stiffness with glycated Hb displaying the highest shear strength. Although the Hb molecule has been classified as a globular protein due to its chemical composition, from our structural analysis, the Hb molecule demonstrates a strikingly anisotropic stiffness behavior as evidenced by its strength in one direction twice larger than the other two directions. We have estimated the different components of the potential energy change during compression and observed that the Coulombic interaction is the main energy component responsible for the material stiffness of hemoglobin. The RMSD results pertaining to the Hb molecules under compression indicate that the active participation of the *heme* protein evolves later in the process, and clearly shows that hemoglobin possesses a soft shell and a rigid core. Thus, from a mechanical behavior standpoint, we conclude that hemoglobin is an anisotropic material with a stiff core and a contiguous soft shell. From our solvated studies and hydrogen bond analysis, we have found that the hemoglobin molecules possess additional strength by creating hydrogen bonds in the presence of water. The mechanical models that were developed in this study can be utilized to develop mesoscopic models that can be employed in multiscale simulations. This study can serve as a basis for the multiscale model development of erythrocytes, and is expected to provide insights on the development of mechanical models of erythrocytes in the area of thrombosis and hemostasis.

## Acknowledgements

Research reported in this publication was supported by the National Heart, Lung, and Blood Institute of the National Institutes of Health under Award Number K01HL115486. The content is solely the responsibility of the authors and does not necessarily represent the official views of the National Institutes of Health. This study was also supported in part by resources and technical expertise from the Georgia Advanced Computing Resource Center, a partnership between the University of Georgia's Office of the Vice President for Research and Office of the Vice President for Information Technology.